\documentclass[lettersize,journal]{IEEEtran}

\usepackage{cite}
\usepackage{amsmath,amssymb,amsfonts}
\usepackage{algorithmic}
\usepackage{graphicx}
\usepackage{textcomp}
\usepackage{xcolor}
\usepackage{booktabs}
\usepackage{url}
\usepackage{float}
\usepackage{placeins}
\usepackage{lettrine}
\usepackage[none]{hyphenat}

\usepackage[hidelinks]{hyperref}
\usepackage{orcidlink}
\usepackage{tikz}  

\newcommand{\ieeecopyrighttext}{%
  \fontsize{6.5}{7.5}\selectfont
  ©\,2026 IEEE. All rights reserved, including rights for text and data
  mining and training of artificial intelligence and similar technologies.
  Personal use is permitted, but republication/redistribution requires
  IEEE permission. See
  \url{https://www.ieee.org/publications/rights/index.html}
  for more information.%
}

\newcommand{\ieeecopyrightnotice}{%
  \begin{tikzpicture}[remember picture,overlay]
    \node[anchor=south,yshift=7pt] at (current page.south) {%
      \parbox{0.92\textwidth}{%
        \centering
        \ieeecopyrighttext
      }%
    };
  \end{tikzpicture}%
}

\newcommand{\ieeeacceptednotice}{%
  \begin{tikzpicture}[remember picture,overlay]
    \node[anchor=north,yshift=-20pt] at (current page.north) {%
      \parbox{0.92\textwidth}{%
        \centering
        \fontsize{6.5}{7.5}\selectfont
        This article has been accepted for publication in
        \emph{IEEE Embedded Systems Letters}. This is the author's version
        which has not been fully edited and content may change prior to final
        publication. Citation information:
        DOI 10.1109/LES.2026.3724013%
      }%
    };
  \end{tikzpicture}%
}

\AddToHook{shipout/foreground}{%
  \ieeeacceptednotice
  \ieeecopyrightnotice
}

\DeclareMathOperator*{\argmax}{arg\,max}

\graphicspath{{Figures/}}

\def\BibTeX{%
  {\rm B\kern-.05em{\sc i\kern-.025em b}\kern-.08em
  T\kern-.1667em\lower.7ex\hbox{E}\kern-.125emX}%
}

\begin{document}

\title{MeanField Surrogate Modeling for Scalable Runtime Scheduling of
Concurrent Heterogeneous AI Inference on Shared GPUs}

\author{%
Youssef Ennouri\,\orcidlink{0009-0005-6741-9660}
and
Soonhoi Ha\,\orcidlink{0000-0001-7472-9142},
~\IEEEmembership{Fellow,~IEEE}%

\thanks{Corresponding author: Soonhoi Ha.
This work was supported by the National Research Foundation of Korea
(NRF), funded by the Korean government (MSIT), under Grant
RS-2026-25483479, with research facilities provided by the Institute of
Computer Technology (ICT), Seoul National University.
The authors are with the Department of Computer Science and Engineering,
Seoul National University, Seoul 08826, Republic of Korea
(e-mail: youssefennouri@snu.ac.kr; sha@snu.ac.kr).}
}

\maketitle

\begin{abstract}
Deploying heterogeneous AI models concurrently on a shared GPU introduces
resource contention that complicates runtime scheduling. While surrogate models
avoid costly online benchmarking, their profiling requirements typically grow
combinatorially with the number of co-running models, limiting scalability. We
propose a MeanField surrogate that predicts per-model performance from local
configuration and aggregate GPU state rather than explicitly modeling all joint
interactions. Experiments on concurrent LLM and vision workloads across
$N \in \{2,3,4,5,6\}$ show high predictive accuracy ($R^2 \approx 0.96$) with
an empirical sample budget that grows approximately linearly in $N$, in contrast
to the combinatorial cost of fully joint profiling. Integrated into a genetic algorithm scheduler, the surrogate scales to a
$N=5$ problem with $78{,}732$ feasible joint configurations, remaining within
$0.10\%$ of the exhaustive search with zero SLA violations across eight dynamic
workload scenarios, while complete online GA decisions take 26 ms median, about $5\times$ faster
than exhaustive surrogate search.
\end{abstract}


\section{Introduction}
\lettrine[lines=2, findent=2pt, nindent=0pt]{M}{}odern inference deployments increasingly co-locate heterogeneous AI workloads such as large language models (LLMs) and vision models on shared GPUs to improve utilization.
Concurrent execution, however, induces non-linear interference over shared resources, including streaming-multiprocessor (SM) occupancy, video memory (VRAM) capacity, and memory bandwidth. This interference depends jointly on model variants, runtime parameters, and workload intensity, making analytical modeling impractical.

Efficient scheduling therefore requires a fast surrogate that predicts
concurrent performance without live benchmarking. While established serving
systems like Nexus~\cite{shen2019nexus}, Clipper~\cite{crankshaw2017clipper}, and
Clockwork~\cite{gujarati2020serving} improve utilization through batching,
placement, and allocation, their scope is limited to homogeneous or isolated workloads. This limitation, along with
interference-aware GPU-sharing systems such as Orion~\cite{orion2024gpu},
ElasticRoom~\cite{elasticroom2024}, Miriam~\cite{miriam2023elastic}, and
automated runtime-aware scheduling~\cite{yu2021runtime} underscores the need for
accurate performance prediction under co-execution.

Scalability, however, remains the obstacle: training a surrogate on joint
$N$-model configurations needs a profiling budget that grows combinatorially in
$N$, and existing surrogate-based approaches~\cite{han2025dvfsgpu} do not address
this for heterogeneous GPU inference.

To address this limitation, we propose a MeanField surrogate formulation with
approximately linear empirical sample-complexity growth in $N$ and integrate it into a practical runtime scheduler.

This paper makes three main contributions:

\begin{itemize}
\item We introduce the MeanField surrogate, a scalable, contention-aware predictor based on local model configurations and
aggregate GPU state.
\item Under a stratified sampling protocol, we empirically show that the
profiling budget required to reach $R^2 \ge 0.95$ grows approximately linearly
over $N \in \{2,\ldots,6\}$, with $n^* \approx 20N$.
\item We develop a runtime scheduler by integrating the MeanField surrogate into a genetic algorithm (GA)-based framework, which is subsequently validated at $N=5$ under dynamic workload scenarios.
\end{itemize}

\begin{figure}[H]
\centering
\includegraphics[width=1\columnwidth]{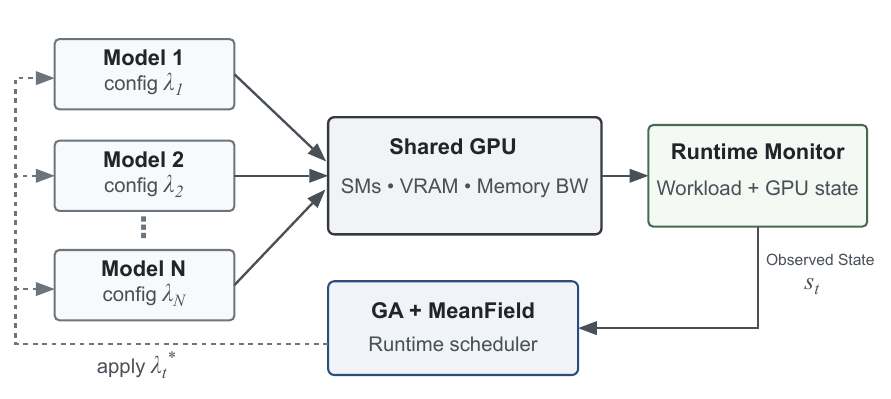}
\caption{The runtime monitor samples the observed state $s_t$, which combines workload
and aggregate GPU information. The MeanField surrogate guides the GA scheduler
in selecting the joint configuration $\lambda_t^\star$ under SLA and VRAM
constraints.}
\label{fig:architecture}
\end{figure}

\section{Runtime Scheduling Model}
\label{sec:formulation}

The system architecture is illustrated in Fig.~\ref{fig:architecture}. Because
exhaustive online benchmarking is infeasible, the scheduler selects
configurations for the co-executing models using a lightweight runtime surrogate
to predict their performance.

To formalize the scheduling problem, we consider $N$ heterogeneous AI models
concurrently executing on a shared GPU. The formulation is agnostic to the
specific application types and applies to any heterogeneous workloads exposing a
discrete configuration space; in this work it is instantiated and evaluated on
two workload families: LLMs based on the Qwen2.5 series and vision models based
on the YOLO11 family.

Each model exposes a discrete configuration space
$\Lambda_i = V_i \times P_i$, where $V_i$ denotes model variants and $P_i$
runtime parameters. For the LLM, $V_i = \{\text{FP16}, \text{AWQ}\}$ (16-bit
floating-point and Activation-aware Weight Quantization variants, respectively)
and $P_i = \{\text{request\_concurrency}\}$ with values in $\{1,2,4\}$; for
YOLO11, $V_i = \{\text{n}, \text{s}, \text{m}\}$ and
$P_i = \{\text{skip\_rate}, \text{imgsz}\}$ with values in $\{1,2,4\}$ and
$\{320,480,640\}$ respectively.
Here, request concurrency controls the number of parallel LLM
requests and therefore affects KV-cache pressure and SM
occupancy.

The scheduler must select one configuration for each model, forming the joint
configuration space:
\begin{equation}
\Lambda = \prod_{i=1}^{N} \Lambda_i
\end{equation}

At each scheduling step, the runtime system observes the state
\begin{equation}
s_t = (\rho_t, m_t, u_t)
\end{equation}
where $\rho_t$ denotes the workload level, $m_t$ the current VRAM usage,
and $u_t$ the aggregate GPU utilization.

The scheduler selects $\lambda^* = \argmax_{\lambda \in \Lambda} J(\lambda, s_t)$,
where the utility $J$ rewards throughput-weighted quality while penalizing
unnecessary reconfigurations and SLA violations:
\begin{equation}
J =
\sum_i \omega_i f_i Q_i
-\alpha \sum_i \eta_i c_i^{\mathrm{rel}}
-\beta \sum_i \max(0,\tau_i+\delta-f_i)^2
\label{eq:objective}
\end{equation}

The first term $\sum_i \omega_i f_i Q_i$ is a throughput-weighted quality
score: a configuration is rewarded only when it is both fast and accurate. The
normalized throughput $f_i \in [0,1]$ uses per-variant min--max scaling,
$f_i = (\hat{f}_i - f^{\min}_v)/(f^{\max}_v - f^{\min}_v)$, where $\hat{f}_i$ is
the raw predicted throughput (tokens/s for the LLM, FPS for vision) and
$f^{\min}_v, f^{\max}_v$ are the extrema observed for variant $v$ during
profiling. Per-variant rather than global normalization is essential: a faster
variant (AWQ) would otherwise dominate the scale and render every configuration
of a slower one (FP16) effectively infeasible. The quality term $Q_i$ captures
variant accuracy ($Q_{\text{LLM}} = \{1.000, 0.979\}$ for FP16/AWQ from relative
perplexity; $Q_{\text{VIS}} = \{0.73, 0.96, 1.00\}$ for YOLO11n/s/m from mean
Average Precision, mAP50-95), scaled so the best variant scores 1.

The priority weights $\omega_i$ are operator-defined and encode
application-specific importance; in our experiments all models are given equal
priority $\omega_i=1$ unless a scenario explicitly elevates one. They may
further be adapted online as
$\omega_i=\bar{\omega}_i+\gamma\max(0,\tau_i-\bar{f}_i)$, with $\gamma=2.0$.

Reconfiguration overhead is modeled through
$\eta_i = \mathbf{1}[v_i^{(t)} \neq v_i^{(t-1)}]$, which indicates whether
model $i$ changes variant and therefore incurs an amortized reload cost
$c_i^{\text{rel}}$ ($c_{\text{LLM}} = 30\,\text{s}$,
$c_{\text{VIS}} = 5\,\text{s}$). The coefficient $\alpha = 1/60 \approx 0.017$ is
the inverse of a 60-second amortization horizon, so a reload is accepted only
when its predicted gain over the next 60\,s exceeds the reload cost itself.

SLA compliance is enforced through a hard per-model throughput floor $\tau_i$.
In the $N=5$ evaluation, $\tau_i$ is set to the empirical P25 of each model's
normalized performance distribution. The quadratic penalty uses a safety margin
$\delta=0.05$ above this floor, discouraging configurations that remain feasible
but operate too close to the SLA boundary. We set $\beta=5.0$.

Hard constraints enforce both VRAM feasibility and minimum throughput
requirements:
\begin{equation}
\sum_i r_i^{\text{VRAM}} \le R, \quad f_i \ge \tau_i
\end{equation}

VRAM-infeasible configurations are removed when constructing $\Lambda$.
Candidates violating any throughput floor receive $J=-\infty$; infeasible
offspring are repaired by replacement with a randomly sampled feasible
individual.

\section{MeanField Surrogate for Scalable Scheduling}
\label{sec:surrogate}

\subsection{Offline Profiling}

Experiments are conducted on a single NVIDIA RTX 3090 (24 GB GDDR6X) running
CUDA 12.6 and Ubuntu 20.04. LLM workloads are served through
vLLM~\cite{kwon2023vllm} with FP16 and AWQ variants. Vision workloads use the
Ultralytics engine.

Offline profiling systematically explores the configuration space to generate
training data for surrogate learning. The baseline $N=2$ configuration space
contains 162 concurrent configurations (6 LLM $\times$ 27 Vision), evaluated
under four workload regimes (low, medium, high, burst), yielding 648 profiling
measurements. For larger systems ($N>2$), additional LLM and vision models were instantiated and profiled under the same workload regimes.

Each cell is measured 10 times (two warmup passes discarded) under fixed GPU
clocks, thermal stabilization below 70\,$^{\circ}$C, and barrier-synchronized
launches.

A profiling sample denotes one sampled joint configuration evaluated under one
workload regime. The ten repeated measurements are averaged into a single
training row and are not counted as distinct samples. For $N>2$, joint
configurations are sampled stratified by model type, variant, and workload
level rather than enumerating the full Cartesian product.

The interference factor for model $i$ is defined as:
\begin{equation}
\Delta_i = f_i^{\text{isolated}} - f_i^{\text{concurrent}}
\end{equation}

Profiling reveals substantial interference: AWQ LLM variants lose 8.4\%
throughput on average under co-execution, while vision models lose 17--24\% FPS
depending on model size. Analysis of variance
(ANOVA)~\cite{montgomery2017design} indicates workload intensity as the dominant
factor for vision degradation ($p<0.01$), whereas the vision-model skip rate
dominates LLM slowdown by reducing shared SM and memory-bandwidth contention.

These results suggest that interference is primarily driven by aggregate
resource pressure rather than by specific model pairings, motivating the
MeanField approximation proposed in this work.

\subsection{MeanField Surrogate}

The name follows the mean-field approximation of statistical physics, in which
the many-body interactions acting on a particle are replaced by a single
aggregate field rather than modeled pairwise. By analogy, each model is
predicted to respond not to the identities of its co-runners but to an aggregate
field, the observed GPU state, that summarizes their combined resource pressure.
The surrogate therefore trains one lightweight predictor per model from the
model's local configuration and aggregate GPU state:
\begin{equation}
\hat f_i = g_i(\lambda_i, m_t, u_t, \rho_t).
\end{equation}

The observed state affects scheduling through the surrogate prediction
$\hat f_i=g_i(\lambda_i,s_t)$: raw throughput $\hat f_i$ is normalized into
$f_i$, which then enters both the throughput-quality reward and the SLA
penalty in Eq.~(3).

During scheduling, $(m_t,u_t,\rho_t)$ are observed context variables obtained
from runtime monitoring or simulation traces and are held fixed during GA
search. The surrogate thus ranks candidate configurations conditioned on the current workload state rather than predicting future GPU-state evolution.

Our measurements support this approximation: aggregate utilization, memory
pressure, and workload level provide sufficient predictive information to retain
high accuracy across the evaluated model mixtures. The
surrogate can therefore be trained directly from concurrent $N$-model
executions, and its profiling cost grows as $O(N \cdot k)$ in the number $k$ of
sampled workload states rather than combinatorially.

Each predictor $g_i$ is a 64--32 MLP with BatchNorm, ReLU,
Dropout$(0.1)$, and Sigmoid output. Categorical features are ordinally
encoded and targets are min--max normalized per variant. This architecture
was selected by ablation at $N=3$ and $n=80$, providing the best
accuracy--latency trade-off and sub-millisecond local inference
(Table~\ref{tab:surrogates}).

\subsection{Scalability Challenge}
\label{sec:scalability}

The main limitation of direct surrogate modeling is the growth of the joint
configuration space. A fully joint predictor requires profiling complexity of
$O(k^N)$, where $k$ denotes the number of configurations per model, and its
input dimensionality grows with $N$. The Joint MLP baseline in
Table~\ref{tab:surrogates} therefore serves only as an $N=2$ reference, as both
profiling requirements and input dimensionality become prohibitive beyond small
deployments.

The MeanField formulation avoids this growth: its profiling cost scales as
$O(N \cdot k)$ rather than $O(k^N)$, while prediction accuracy remains high
(Section~\ref{sec:results}).

To characterize this trade-off, the MeanField surrogate is trained from scratch
and evaluated using stratified cross-validation across $N \in \{2,3,4,5,6\}$, measuring
both its asymptotic accuracy and the number of samples required to reach
convergence.

\vspace{2cm}

\section{Experimental Results}
\label{sec:results}

\subsection{Sample-Complexity Scaling}
\label{sec:scaling}

We evaluate surrogate sample efficiency under identical training protocols
across $N \in \{2,3,4,5,6\}$. Performance is measured using the mean $R^2$ across targets, and convergence
is defined as the minimum training set size $n^*$ required for the mean
$R^2$ to reach 0.95. Results are averaged over 25
evaluations (5 random seeds $\times$ 5-fold stratified cross-validation),
capturing both data-sampling and optimization variance.

\begin{figure}[H]
\centering
\includegraphics[width=1.05\columnwidth]{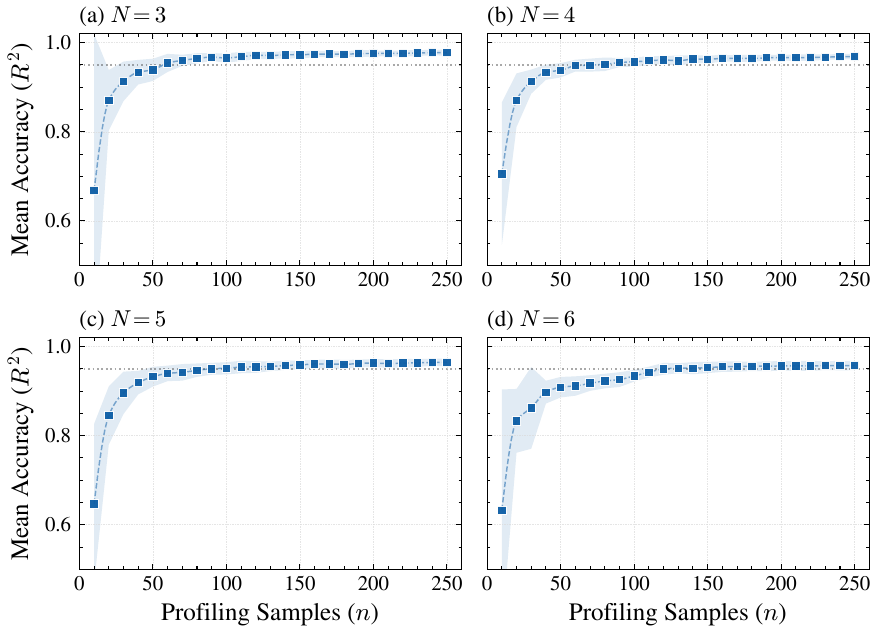}
\caption{Mean $R^2$ vs. training sample count $n$ for the MeanField surrogate
at $N \in \{3,4,5,6\}$. At every $N$, the surrogate reaches the
$R^2 \geq 0.95$ threshold within a modest sample budget, with the required
budget increasing approximately linearly as $N$ grows.}
\label{fig:curvesN3_N6}
\end{figure}

Fig.~\ref{fig:curvesN3_N6} shows the MeanField learning curves under increasing
training budgets. At every value of $N$ the surrogate reaches the target
accuracy with a small number of samples and then plateaus near
$R^2 \approx 0.95$--$0.97$, with low inter-seed and inter-fold variability.

\begin{figure}[H]
\centering
\includegraphics[width=0.90\columnwidth]{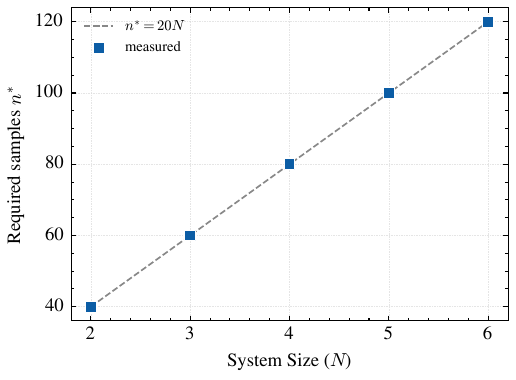}
\caption{Empirical convergence sample count $n^*$ for the MeanField surrogate.
The observed budget grows approximately linearly, $n^* \approx 20N$, over
$N \in \{2,3,4,5,6\}$.}
\label{fig:scaling_law}
\end{figure}

Fig.~\ref{fig:scaling_law} summarizes this behavior through the empirical
convergence sample count $n^*$. MeanField reaches the target accuracy after
approximately (40, 60, 80, 100, 120) samples for $N \in \{2,3,4,5,6\}$,
consistent with a near-linear trend ($n^* \approx 20N$) and in sharp contrast to
the combinatorial $O(k^N)$ cost of a joint surrogate.

\subsection{Prediction Accuracy}

Table~\ref{tab:surrogates} reports the mean $R^2$ of the MeanField surrogate at
full training budget. Accuracy remains high ($R^2 > 0.95$) up to $N=6$ and
stays within a narrow band of the $N=2$ Joint MLP reference, confirming that the
scalable MeanField formulation sacrifices little asymptotic accuracy while
removing the joint surrogate's combinatorial profiling cost.

\begin{table}[H]
\caption{Asymptotic prediction accuracy (Mean $R^2 \pm$ std over 25 evaluations)
of the MeanField surrogate across $N$, with the joint MLP shown as an $N=2$
reference.}
\label{tab:surrogates}
\centering
\begin{tabular}{ccc}
\toprule
\textbf{N} & \textbf{Joint MLP} & \textbf{MeanField} \\
\midrule
2 & $0.9826 \pm 0.002$ & $0.9860 \pm 0.003$ \\
3 & -- & $0.9779 \pm 0.006$ \\
4 & -- & $0.9689 \pm 0.006$ \\
5 & -- & $0.9645 \pm 0.014$ \\
6 & -- & $0.9588 \pm 0.013$ \\
\bottomrule
\end{tabular}
\end{table}

\subsection{Runtime Scheduling Validation}

The MeanField surrogate drives a GA scheduler at \mbox{$N=5$}, over a
VRAM-feasible space of $78{,}732$ configurations spanning three LLMs and two
YOLO-family models. The 7B LLM is AWQ-only due to the 24 GB memory limit, and
per-model SLA floors are set to the empirical P25.

The GA uses 100 candidates, crossover probability 0.7, mutation probability
0.1, tournament selection ($k=3$), elitism, and 30 generations. MeanField 
evaluates all local configurations in a median of 8.5 ms, followed by 17 ms for GA
search, yielding a 26 ms end-to-end decision latency. Exhaustive surrogate
search requires 131 ms, making the GA about $5\times$ faster.

\begin{table}[H]
\centering
\caption{Scheduling results at $N=5$ over eight dynamic scenarios.}
\label{tab:scheduler_n5}
\footnotesize
\setlength{\tabcolsep}{4pt}
\begin{tabular}{lcccc}
\toprule
\textbf{Scheduler} & \textbf{Mean $J$} & \textbf{Avg. Gap} & \textbf{Worst Gap} & \textbf{SLA} \\
          &          & \textbf{(\%)} & \textbf{(\%)} & \textbf{(\%)} \\
\midrule
Exhaustive Search   & 2.744 & 0.00 & 0.00 & 0.00 \\
StaticFull          & 2.743 & 0.05 & 0.41 & 0.00 \\
GA (ours)           & 2.742 & 0.10 & 0.38 & 0.00 \\
StaticBudget ($K=5000$)  & 2.600 & 5.22 & 7.36 & 0.00 \\
StaticBudget ($K=500$)   & 2.443 & 10.51 & 17.16 & 0.00 \\
Random              & 1.428 & 49.13 & 58.71 & 3.45 \\
\bottomrule
\end{tabular}
\end{table}

Table~\ref{tab:scheduler_n5} reports mean utility over eight dynamic scenarios
and three seeds. Exhaustive denotes a per-step argmax over the same
surrogate-predicted landscape, so the reported gap measures search quality only.

We compare against StaticFull, StaticBudgetK, and Random.
StaticFull is an adaptivity ablation with full-space access, whereas
StaticBudgetK selects one fixed configuration from only $K$ offline samples.

The GA remains within $0.10\%$ of exhaustive search with zero SLA violations.
StaticFull is also near-optimal, indicating limited adaptation benefit in these
scenarios. Under realistic calibration budgets, however, StaticBudgetK loses
$10.51\%$ utility at $K=500$ and $5.22\%$ at $K=5000$. The main benefit is
therefore profiling-efficient near-optimal search rather than adaptation alone.

\section{Conclusion}

This paper proposed a MeanField surrogate for scalable runtime scheduling of
concurrent heterogeneous AI workloads on a shared GPU. Across
$N \in \{2,3,4,5,6\}$, it achieved high prediction accuracy
($R^2 \approx 0.96$) with approximately linear empirical sample-complexity
growth. Integrated into a GA scheduler, it scaled to an $N=5$ problem with $78{,}732$
feasible configurations, remaining within $0.10\%$ of exhaustive search with zero SLA
violations across eight dynamic scenarios, with a median end-to-end decision latency of 26 ms.
Future work will extend the LLM workload model to capture prefill/decode
phases and KV-cache occupancy.

\bibliographystyle{IEEEtran}
\bibliography{refs}

\end{document}